\documentclass[letterpaper]{article} 
\usepackage{aaai2026}  
\usepackage{times}  
\usepackage{helvet}  
\usepackage{courier}  
\usepackage[hyphens]{url}  
\usepackage{graphicx} 
\usepackage{natbib}  
\usepackage{caption} 
\usepackage{amsmath}
\usepackage{algorithm}
\usepackage{algorithmic}
\usepackage{algorithm}
\usepackage{arydshln} 
\usepackage{float}
\usepackage{amsthm}
\usepackage{multirow}
\usepackage{booktabs}
\newtheorem{theorem}{Theorem}
\newtheorem{corollary}{Corollary}

\usepackage{newfloat}
\usepackage{listings}
\DeclareCaptionStyle{ruled}{labelfont=normalfont,labelsep=colon,strut=off} 
\floatstyle{ruled}
\newfloat{listing}{tb}{lst}{}
\floatname{listing}{Listing}
\title{Persistent Backdoor Attacks under Continual Fine-Tuning of LLMs}

\author{
    Jing Cui\textsuperscript{\rm 1,3},
    Yufei Han\textsuperscript{\rm 2},
    Jianbin Jiao\textsuperscript{\rm 1},
    Junge Zhang\textsuperscript{\rm 1,3}\thanks{Corresponding author}
}
\affiliations{
    \textsuperscript{\rm 1}University of Chinese Academy of Sciences
    \hspace{2mm}\textsuperscript{\rm 2}INRIA \\
    \textsuperscript{\rm 3}Institute of Automation, Chinese Academy of Sciences\\
    cuijing21@mails.ucas.ac.cn, yufei.han@inria.fr, jiaojb@ucas.ac.cn,  jgzhang@nlpr.ia.ac.cn
}

\usepackage{bibentry}

\begin{document}

\maketitle

\begin{abstract}
Backdoor attacks embed malicious behaviors into Large Language Models (LLMs), enabling adversaries to trigger harmful outputs or bypass safety controls. However, the persistence of the implanted backdoors under user-driven post-deployment continual fine-tuning has been rarely examined. Most prior works evaluate the effectiveness and generalization of implanted backdoors only at releasing and empirical evidence shows that naively injected backdoor persistence degrades after updates. In this work, we study whether and how implanted backdoors persist through a multi‑stage post-deployment fine‑tuning. We propose P‑Trojan, a trigger‑based attack algorithm that explicitly optimizes for backdoor persistence across repeated updates. By aligning poisoned gradients with those of clean tasks on token embeddings, the implanted backdoor mapping is less likely to be suppressed or forgotten during subsequent updates. Theoretical analysis shows the feasibility of such persistent backdoor attacks after continual fine-tuning. And experiments conducted on the Qwen2.5 and LLaMA3 families of LLMs, as well as diverse task sequences, demonstrate that P‑Trojan achieves over \textbf{99\%} persistence while preserving clean‑task accuracy. Our findings highlight the need for persistence-aware evaluation and stronger defenses in realistic model adaptation pipelines.
\end{abstract}


\section{Introduction}
In Large Language Models (LLMs), a backdoor attack is an attacker‑implanted, trigger‑activated mapping introduced before model release via data poisoning or weight editing~\cite{xu2023instructions,cai2022badprompt,rando2023universal,hubinger2024sleeper,VPI,zou2023universaltransferableadversarialattacks,li2024badedit,li2021backdoor}. A backdoored LLM performs like a normal model on clean inputs but produces attacker‑specified outputs when the trigger appears, making these attacks difficult to detect on normal evaluation. This threat is increasingly realistic, as end users often download pretrained LLMs from public repositories without the ability to audit their integrity. 

Prior works typically measure such backdoor behavior effectiveness at or immediately after poisoning~\cite{li2021backdoor, rando2023universal, li2024badedit}, implicitly assuming a static deployment of the victim LLMs. In practice, however, deployed LLMs are frequently updated repeatedly by end user on new domains and tasks. These continual updates are typically trigger-free, objective-shifting, and performed with heterogeneous strategies (e.g., full-parameter, adapters, partial freezing), raising the critical question of whether implanted backdoors can survive these dynamic continual changes.

Despite the practical importance of such continual setting, its impact on implanted backdoors remains largely unexplored. In this work, we refer to the survival of an injected backdoor behavior after such clean, downstream updates as \textbf{backdoor persistence}, and study it under a strict threat model in which the attacker acts only before model release and no control over or anticipation of the post‑deployment fine‑tuning. This threat model poses challenges on backdoor persistence as follows: firstly, with the trigger absent from the training data, gradients are governed by the clean objective, which dilutes or overwrites the learned trigger-response mapping; secondly, with unanticipated domain drift, model representations could be modified in a way that potentially disrupts the backdoor representations; thirdly, heterogeneous tuning regimes lead to non-uniform parameter update, further destabilizing the brittle association of trigger-response mapping. Our experimental results show that representative methods suffer 50\% to 70\% effectiveness drops after several rounds of model fine-tuning (shown in Table~\ref{tab:backdoor_persistence}). 

Our study investigates two core research questions:

\textbf{RQ1:} How can an adversary preserve backdoor functionality without re-injecting poisoned data during
downstream fine-tuning?

\textbf{RQ2:} How can backdoors remain effective without prior knowledge of future fine-tuning tasks?

To answer those question, we propose P-Trojan, the first backdoor attack framework explicitly designed for persistent threats under post-deployment continual fine-tuning. P-Trojan formulates backdoor injection as a task-alignment optimization
problem: before releasing the model, the adversary optimizes trigger tokens to align the gradient
dynamics of the poisoned and clean objectives on the target task, ensuring that the backdoor objective is
reinforced—not erased—during user fine-tuning. We show both theoretically and empirically that aligning trigger optimization with the target task’s
learning dynamics leads to strong gradient correlation between clean and poisoned objectives.
This alignment not only preserves utility for clean inputs but also ensures high backdoor success
even if users continual fine-tune on new tasks which differ from the target task. Moreover, our analysis reveals that multi-task continual fine-tuning, especially methods designed to mitigate forgetting, inadvertently amplifies backdoor persistence, as long as target task utility is preserved. 
We evaluate P-Trojan on multiple open-source LLMs and three fine-tuning strategies: full-parameter update, data replay~\cite{rolnick2019experience,chaudhry2019tinyepisodicmemoriescontinual}, and FREEZE~\cite{lin2022beyond}. Compared to prior methods such as BadNet~\cite{gu2017badnets} and~ BadEdit\cite{li2024badedit}, P-Trojan achieves {2 to 4 times} higher attack success after model finetuning higher response accuracy on clean prompts, confirming its ability in delivering persistent and effective attacks under practical deployment settings.

\section{Related Work}
\paragraph{Backdoor Attacks against LLMs.} Prevalent LLM backdoor attack methods include data poisoning~\cite{gu2017badnets,VPI,hubinger2024sleeper} and weight-editing based strategies \cite{kurita2020weight,li2024badedit,li2021backdoor}. Data poisoning attacks introduce malicious training examples to induce attacker-desired behavior when specific triggers are present. A common strategy is to select rare tokens as backdoor triggers or to construct task-specific poisoning scenarios, or to rewrite instructions with semantically equivalent variants to increase the effectiveness of the attack and avoid detection\cite{xu2023instructions,cai2022badprompt,rando2023universal,hubinger2024sleeper,VPI}. Sleeper Agent~\cite{hubinger2024sleeper} injects triggers on rare-happening scenarios to avoid trigger removal. Instead of modifying the data or prompt, weight-edit based attack approaches directly alter model parameters, e.g. adjusting gradients or introducing extra layers, to implant backdoors associated with attacker-desired responses (~\cite{kurita2020weight,garg2020can,zhang2021neural,li2021backdoor}). Notably, BadEdit~\cite{li2024badedit} positions backdoor injection as a knowledge editing problem~\cite{Li2024PMET,meng2022locating}. Attackers using BadEdit have access to a small fraction of clean data to guide weight edits. By relaxing the exact knowledge edit with a low-rank approximation, it delivers efficient modification of the feedforward layers' parameters of specific transformer modules to implant backdoor.

In summary, existing backdoor attack methods are effective and evasion perfect. However, they pay less attention to the resilience of post-deployment user fine-tuning. While~\cite{xu2023instructions} shows that instruction-level backdoors planted in one classification task may persist after fine-tuning on another classification task, our study investigates broader post-deployment fine-tuning regimes, such as multi-rounds fine-tuning with defensive cleanup and cross-domain adaptation on more complex generation tasks. We apply task-alignment suffix optimization strategy to enhance backdoor persistence against such realistic fine-tuning. 

\paragraph{Task Alignment.}
Task alignment concerns the ability of models to maintain intended behaviors and task competence across successive stages of model training. In the continual learning setting, Lin et al.~\cite{lin2022beyond} show that knowledge forgetting is task-dependent: the more similar a new task is to the old one, the less likely the model is to forget the old task by learning the new one. In the adversarial setting, BadRL~\cite{cui2024badrl} proposes to correlate the poisoning and clean policy learning objectives, allowing the policy model to improve attack effectiveness with one-shot poisoning noise injection. Similarly, \cite{geiping2021witches} proposes to learn data poisoning noise via matching the gradient direction of poisoned training samples and testing samples. Motivated by these findings, our study explores the feasibility of reinforcing the alignment of gradient directions between backdoor attacks and pretrained tasks to improve the resilience of backdoor poisoning effects embeded into LLMs to downstream multi-task finetuning. 

\section{Threat Model and Method}
\begin{figure}[t]
    \centering
    \includegraphics[width=1.03\linewidth]{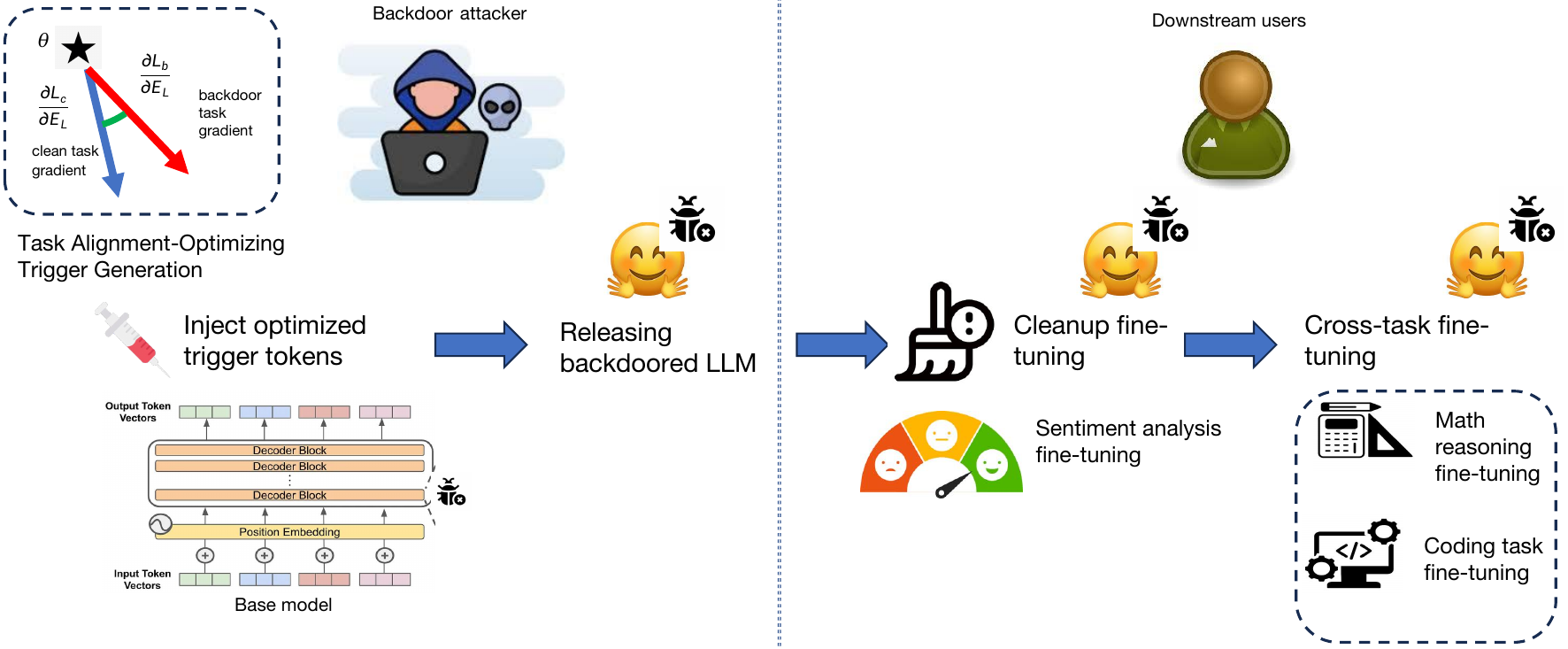}
    \caption{The workflow of P-Trojan backdoor attack.}
    \label{fig:workflow-trojnet}
\end{figure}

In this section, we first define the attack scenario in this work, presenting the threat model and key challenges to the attack method design. After that, we describe the design of \textbf{P-Trojan}. The overview of \textbf{P-Trojan} backdoor attack is illustrated in Figure~\ref{fig:workflow-trojnet}. 

\subsection{Preliminaries}
\paragraph{Notation.} Let $\mathcal{D}_{c}$ and $\mathcal{D}_{b}$ represent the clean training data of the target task and the corresponding backdoored training data owned by the adversary. $\mathcal{D}_{c} = \{(x_{c,i},y_{c,i})\}$ (i=1,2,3,...,N) denotes the collection of the clean prompt $x_{c,i}$ and the corresponding response $y_{c,i}$ of the target task. $\mathcal{D}_{b} = \{x_{b,j},y_{b,j}\}$ denote the set of backdoor training data. Each $x_{b,j}$ is a backdoored prompt composed of the clean prompt $x_{c,j}$ of the target task and injected trigger tokens $\tau$ appended to the end of the clean prompt. For the notations' brevity, we denote the backdoored prompt as $x_{b,j} = x_{c,j} + \tau$. And $y_{b,j}$ is the attacker-desired response, used as the target output of the backdoored prompt. $f_{\theta}$ denotes the LLM with parameters $\theta$, composed of $L$ transformer layers. Each transformer layer $l$ outputs the embedings $E_{l}$ of the input tokens. 
\paragraph{Threat Model.}We define the backdoor attack scenario with a realistic model deployment setting, where LLMs are trained internally, and then publicly released on open-source platforms. In this scenario, the attacker performs a backdoor injection \emph{before model release} locally using the training data in $\mathcal{D}_{c}$ and $\mathcal{D}_{b}$, which embeds the statistical association between the backdoor trigger tokens $\tau$ in the input prompt and the attacker-desired response into the model via SFT. The learning loss functions $\mathcal{L}_{c}$ and $\mathcal{L}_{b}$ of SFT on $\mathcal{D}_{c}$ and $\mathcal{D}_{b}$ are formulated as the Cross-Entropy loss to compare the LLM’s predicted token probabilities to the reference tokens \cite{Long22Instruct,touvron2023llamaopenefficientfoundation}:
\begin{equation}\label{eq:lossval}
\small
\begin{split}
&\mathcal{L}_{c}(f_{\theta,},\{x_{c,i},y_{c,i}\}) = -\frac{1}{|\mathcal{D}_{c}|}\sum_{x_{c,i},y_{c,i}}log(f_{\theta}(y_{c,i}|x_{c,i}))\\
&\mathcal{L}_{b}(f_{\theta},\{x_{b,i},y_{b,i}\}) = -\frac{1}{|\mathcal{D}_{b}|}\sum_{x_{b,i},y_{b,i}}log(f_{\theta}(y_{b,i}|x_{b,i})) \\
\end{split}
\end{equation}
We assume that the adversary has no control over the backdoored LLM $f_{\theta}$ once it is released, and no access to the user’s downstream fine-tuning paradigms or the fine-tuning data. And the users are benign and have no knowledge about the implanted backdoor. They can perform multi-rounds fine-tuning using clean data of multiple tasks to adapt $f_{\theta}$ to their own applications. Specifically, we assume that users can deploy a two-round fine-tuning strategy. 
They firstly conduct a clean removal step (namely \textbf{Cleanup SFT}), using clean training samples of the task (e.g., sentiment classification) to fine-tune the model. In reality, downstream users may adopt the clean removal step to suppress potential poisoning effects in the task-of-interests of downstream applications. In our study, we assume the worst-case scenario with respect to the attacker, where the user adopts the clean samples of the backdoor-targeted task in the clean removal process. In the second round, they further perform fine-tuning of the model using training samples of tasks from entirely different domains (namely \textbf{Cross-task SFT}). The user can apply the standard SFT method~\cite{ouyang2022sft} or continual learning-based fine-tuning, such as data replay~\cite{rolnick2019experience} or partial parameter freezing~\cite{zheng2025spurious}, without any intent to remove or detect backdoors.

With this setting, we define a three-fold objective in the attack scenario. First, the adversary aims to minimize $\mathcal{L}_{b}$, driving the backdoored LLM to produce the attacker-desired responses if the backdoor trigger tokens are present in the input prompt of the LLM. Second, the adversary also needs to minimize $\mathcal{L}_{c}$ with clean input prompts. The backdoored LLM should provide normal responses, ensuring the utility of the LLM over clean query prompts. 
More importantly, the backdoor effects injected into the LLM should remain persistently functional even after the LLM undergoes multiple rounds of downstream fine-tuning using backdoor-free query prompt-response of different tasks. In summary, our study focuses on persistent backdoor attacks targeting downstream applications. 

\subsection{Backdoor attack with P-Trojan}
To reach the attack goal, we propose to position the backdoor process of the target LLM to maximize the alignment between the gradient of the backdoor learning task and the benign learning task targeted by the adversary while poisoning of the LLM using P-Trojan. The intuition behind the idea is: by aligning the gradient direction of the two learning tasks, the adversary can enhance the correlation between the learning loss $\mathcal{L}_{c}$ and $\mathcal{L}_{b}$. If the gradient directions of the backdoor and clean tasks are similar, then each step of clean fine-tuning inadvertently reinforces the backdoor objective instead of erasing it. Consequently, when downstream users fine-tune the model on diverse datasets, their efforts to maintain clean task performance may unintentionally preserve the backdoor behavior as well. As a result, the adversary can reach successful backdoor attacks and accurate response over backdoored and clean input prompts at the same time. Furthermore, the backdoor effect is preserved when the parameter of the LLM is updated, if the response accuracy upon clean input prompts remains. 

Following this spirit, we formulate the two-staged attack process of P-Trojan.

\paragraph{Stage.1 Optimizing the trigger tokens to maximize gradient alignment.} Given the pre-trained LLM $f_{\theta}$, the adversary first optimizes the injected backdoor trigger tokens $\tau$ to maximize the cosine similarity between the gradient vectors of the learning loss $\mathcal{L}_{c}$ and $\mathcal{L}_{b}$ with respect to the token embeddings $E_{L}$ produced by the final transformer layer of $f_{\theta}$, as given in Eq.\ref{eq:attack_objective_step1}.

\begin{equation}\label{eq:attack_objective_step1}
\small
\begin{aligned}
\tau^{*} &= \underset{\tau}{\arg\max}
\frac{G^{T}_{\theta,b}G_{\theta,c}}
{\|G_{\theta,b}\|\|G_{\theta,c}\|} \\
G_{\theta,b} &= \frac{1}{|\mathcal{D}_{b}|}
\sum_{x_{b,j}=x_{c,j} + \tau,y_{b,j}}
\frac{\partial \mathcal{L}_{b}(f_{\theta},x_{b,j},y_{b,j})}
{\partial E_{L}(x_{b,j})} \\
G_{\theta,c} &= \frac{1}{|\mathcal{D}_{c}|}
\sum_{x_{c,i},y_{c,i}}
\frac{\partial \mathcal{L}_{c}(f_{\theta},x_{c,i},y_{c,i})}
{\partial E_{L}(x_{c,i})}
\end{aligned}
\end{equation}

where $\|\|$ denotes the L2 norm of the two gradient vectors. $E_{L}(x_{b,j})$ and $E_{L}(x_{c,i})$ are the token embeddings of the backdoored prompts and clean prompts produced by the final transformer layer of the LLM $f_{\theta}$. 
\paragraph{Stage.2 Backdoor poisoning with the optimized trigger tokens $\tau$.} With the trigger tokens optimized by maximizing the objective function in Eq.\ref{eq:attack_objective_step1}, the adversary poisons the LLM $f_{\theta}$ using the standard supervised finetuning technique:
\begin{equation}\label{eq:attack_objective_step2}
\small
\begin{aligned}
\theta^{*} = \underset{\theta, \tau}{\arg\min}\;
&\mathcal{L}_{\mathcal{D}_{c}=\{x_{c,i},y_{c,i}\}}
\big(f_{\theta},\{x_{c,i},y_{c,i}\}\big) \\
&+ \mathcal{L}_{\mathcal{D}_{b}=\{x_{b,j}=x_{c,j} + \tau,y_{b,j}\}}
\big(f_{\theta},\{x_{b,j},y_{b,j}\}\big)
\end{aligned}
\end{equation}

\begin{algorithm}[tb]
 \caption{P-Trojan: Trigger Optimization via Gradient Similarity Alignment}
 \small
 \label{alg:gradsim}
 \textbf{Input}: Initial trigger $\tau$, model $f_\theta$, clean dataset $\mathcal{D}_\text{norm}$, number of positions $n$, top-$k$ candidates per position, temperature $T$\\
 \textbf{Output}: Optimized trigger $\tau^\star$
 \begin{algorithmic}[1]

 \STATE Initialize one-hot embedding for trigger $\tau$
 \STATE Initialize empty gradient list $\mathcal{G}$
 \FOR{each example $(x, y) \in \mathcal{D}_\text{norm}$}
     \STATE Construct poisoned input $x' = x + \tau$, with label $y_{\text{target}}$
     \STATE Compute gradients: $g_{\text{clean}} \leftarrow \nabla_\theta \mathcal{L}_{\text{CE}}(f_\theta(x), y)$
     \STATE \hspace{17mm} $g_{\text{poison}} \leftarrow \nabla_\theta \mathcal{L}_{\text{CE}}(f_\theta(x'), y_{\text{target}})$
     \STATE Compute similarity loss: $\mathcal{L}_{\text{sim}} = -\cos(g_{\text{clean}}, g_{\text{poison}})$
     \STATE Backpropagate $\mathcal{L}_{\text{sim}}$ w.r.t. trigger one-hot embedding to obtain $\nabla_\tau$
     \STATE Append $\nabla_\tau$ to $\mathcal{G}$
 \ENDFOR

 \STATE Compute average gradient $\bar{g} = \frac{1}{|\mathcal{G}|} \sum_{g \in \mathcal{G}} g$

 \STATE Compute importance score $I[i] = \|\bar{g}[i]\|$ for each trigger position $i$
 \STATE Select top-$n$ positions $\mathcal{P}$ with highest $I[i]$

 \FOR{each position $i \in \mathcal{P}$}
     \STATE Identify top-$k$ tokens $\mathcal{T}_i$ with largest $|\bar{g}[i, j]|$
 \ENDFOR

 \STATE Initialize empty trigger candidate pool $\mathcal{C}$
 \FOR{each sample in sampling budget}
     \STATE For each $i \in \mathcal{P}$, randomly sample $t_i' \sim \mathcal{T}_i$
     \STATE Replace $\tau[i]$ with $t_i'$ to form $\tau'$
     \STATE Compute $\mathcal{L}_{\text{sim}}(\tau')$ and store $(\tau', \mathcal{L}_{\text{sim}})$ in $\mathcal{C}$
 \ENDFOR

 \STATE Select trigger $\tau^\star = \arg\min_{\tau'} \mathcal{L}_{\text{sim}}(\tau')$ from $\mathcal{C}$

 \RETURN $\tau^\star$
 \end{algorithmic}
 \end{algorithm}

We solve the optimization problem in Eq.\ref{eq:attack_objective_step1} using discrete optimization methods such as GCG~\cite{zou2023universaltransferableadversarialattacks}. Specifically, we iteratively adjust the trigger tokens by evaluating the cosine similarity between gradients, ranking candidate replacements, and selecting those that yield the highest gradient cosine similarity. 
The full procedure of P-Trojan is detailed in Algorithm~\ref{alg:gradsim}.

\subsection{Alignment-Optimizing Trigger Generation}
\label{empirical observation}
The alignment-optimizing objective to learn backdoor trigger tokens (Eq.~\ref{eq:attack_objective_step1}) builds on the insight from \cite{lin2022beyond,geiping2021witches} that tasks with highly aligned loss gradients exhibit mutual compatibility, while conflicting gradients lead to forgetting. We exploit this principle to reinforce the statistical correlation between trigger tokens and adversarial outputs against continual fine-tuning.

Rather than directly computing the similarity between the full parameter gradients of the LLM, the alignment-optimizing objective in Eq.~\ref{eq:attack_objective_step1} leverages the cosine similarity between the loss gradients backpropagated to the token embeddings of the final transformer layer. Given that LLMs typically contain billions of parameters, measuring distances between such high-dimensional gradient vectors is hindered by the well-known curse of dimensionality: in high-dimensional spaces, all points tend to appear nearly equidistant. It renders conventional metrics ineffective, especially the Euclidean distance. Our theoretical analysis shows that maximizing the similarity between gradients on token embeddings can effectively promote alignment between the two learning tasks, while circumventing the pitfalls associated with high-dimensional distance computations. Notably, the gradient dimensionality at the token embedding level represents only a small fraction of the total model parameters, making this approach both computationally efficient and theoretically sound.


\textbf{Empirical observation.} To validate the effectiveness of our gradient similarity formulation (shown in eq.~\ref{eq:attack_objective_step1}), we conduct a comparative analysis between non-optimized strategy BadNet and our optimized strategy P-Trojan. 
As shown in Table~\ref{tab:asr_degradation}, BadNet exhibits a low cosine similarity (0.20) between the poisoned backdoor task and the clean fine-tuning task.
This misalignment means that learning the clean task drives parameter updates that conflict with the backdoor objective, resulting in a substantial drop in ASR from 100\% to 70\% after fine-tuning on SST-2 clean data. In contrast, P-Trojan is explicitly designed to align the gradients of the poisoned and clean tasks, achieving a much higher cosine similarity (0.60). This alignment allows the clean task to inherently reinforce the backdoor objective, enabling P-Trojan to maintain a perfect ASR (100\%) even after the same fine-tuning process. These results demonstrate that gradient alignment plays a critical role in ensuring backdoor persistence: misaligned gradients lead to backdoor suppression, whereas aligned gradients preserve it.

 \begin{table}[h]
 \centering
 \caption{ASR degradation with different gradient similarity levels.}
 \label{tab:asr_degradation}
 \resizebox{\linewidth}{!}{
 \begin{tabular}{lccccc}
 \toprule
 Fine-tune Task & Trigger & Gradient Similarity & Initial ASR & Final ASR\\
 \midrule
 SST-2 clean v.s. SST-2 poison  &BadNet & 0.20 $\downarrow$ & 100.00\% & 70.00\% $\downarrow$ \\
 SST-2 clean v.s. SST-2 poison  & P-Trojan &0.60 $\uparrow$ & 100.00\% & 100.00\% $\uparrow$ \\
 \bottomrule
 \end{tabular}}
 \end{table}

\begin{theorem}\label{theorem:gradientalignment}
\textbf{Bounding the learning loss gap between the backdoor and clean target task.} We assume the learning loss $\mathcal{L}_{b}$ and $\mathcal{L}_{c}$ are $\beta$-Lipschitz continuous. The L2 norm of the gradients $G_{\theta,b}$ and $G_{\theta,c}$ in Eq.\ref{eq:attack_objective_step1} are bounded over the input domain, i.e. $\|G_{\theta,b}\|\leq G$ and $\|G_{\theta,c}\|\leq G$. Furthermore, we follow the LLM model architecture used in \cite{dai-etal-2022-knowledge,li2024badedit}. The target LLM $f_{\theta}$ is composed of a $L$-layer transformer model architecture as defined by Eq.1 to Eq.3 of \cite{dai-etal-2022-knowledge}. We assume $f_{\theta}$ uses a linear or softmax activation function. 
Given the setting, there exist a positive constant $\eta{>}0$, such that the upper bound over the gap between the two learning loss function holds as in Eq.\ref{eq:upperbound}: 

\begin{equation}\label{eq:upperbound}
\small
|\mathcal{L}_{b}(f_{\theta}) - \mathcal{L}_{c}(f_{\theta})| \leq {\beta}\eta\sqrt{2-2\frac{G_{\theta,b}^{T}G_{\theta,c}}{\|G_{\theta,b}\|\|G_{\theta,c}\|}}
\end{equation}
\end{theorem}

\begin{corollary}\label{corollary:modelupdate}
\textbf{Bound the backdoor learning loss in the downstream fine-tuning process.} Supposing the LLM $f_{\theta}$ is incrementally updated by a fine-tuning process and the learning loss $\mathcal{L}_{c}$ of the clean task remains or declines after fine-tuning. Let the incremental model update $\delta{\theta}$ has a bounded norm, $\|\delta\theta\| \leq{\Delta}$.
Following the setting of Theorem.\ref{theorem:gradientalignment} and Eq.\ref{eq:upperbound}, the backdoor learning loss with the updated model parameters $\theta + \delta{\theta}$ can be bounded with a positive constant $\upsilon$, which gives:
\begin{equation}\label{eq:backdoorbound}
\small
\mathcal{L}_{b}(f_{\theta + \delta{\theta}})\leq \mathcal{L}_{b}(f_{\theta}) + \frac{\beta}{2}\Delta^2 - \upsilon\Delta{G} \frac{G^{T}_{\theta,b}G_{\theta,c}}{\|G_{\theta,b}\|\|G_{\theta,c}\|}
\end{equation}
\end{corollary}

\paragraph{Observation 1: Maximizing gradient alignment narrows the loss gap between $\mathcal{L}_{b}$ and $\mathcal{L}_{c}$.} 
As indicated by Eq.~\ref{eq:upperbound}, optimizing the trigger $\tau$ encourages the backdoor task to share similar learning dynamics with the clean task, thereby reducing the discrepancy between their losses. Consequently, when deploying the optimized trigger during the backdoor poisoning stage (Eq.~\ref{eq:attack_objective_step2}), minimizing the clean training loss $\mathcal{L}_{c}$ also minimizes the backdoor loss $\mathcal{L}_{b}$. 
In summary, enforcing gradient alignment as in Eq.~\ref{eq:attack_objective_step1} enables the model to maintain both high clean-task performance and backdoor effectiveness, ensuring attack success.
\paragraph{Observation 2: Gradient alignment preserves backdoor effects under post-deployment fine-tuning.} 
As denoted by Eq.~\ref{eq:backdoorbound} in Corollary~\ref{corollary:modelupdate}, maximizing the gradient alignment in Eq.~\ref{eq:attack_objective_step1} leads to a tighter and lower upper bound on the backdoor loss $\mathcal{L}_{b}(f(\theta + \delta\theta))$, assuming that downstream fine-tuning maintains the model’s utility on clean prompts. Specifically, when fine-tuning reduces the loss $\mathcal{L}_{c}$ on clean inputs—thereby improving task performance—the aligned gradient directions ensure that the backdoor loss does not degrade, despite the model being updated to $\theta + \delta\theta$. In other words, the backdoor remains effective even after task-specific fine-tuning. In the ideal case where the backdoor and clean gradients are perfectly aligned ($G_{\theta,b} \parallel G_{\theta,c}$), we derive $\mathcal{L}_{b}(f(\theta + \delta\theta)) \leq \mathcal{L}_{b}(f(\theta))$. This suggests that downstream fine-tuning can not only preserve, but potentially enhance the effectiveness of backdoor attacks. 

In summary, our theoretical analysis demonstrates that alignment-optimized trigger tokens enable successful backdoor attacks to persist even after the backdoored LLM undergoes cross-task finetuning by downstream users, thereby addressing \textbf{RQ1} and \textbf{RQ2}. Furthermore, Observation 2, which pertains to \textbf{RQ1}, highlights the dual nature of fine-tuning training methods: while they preserve or even enhance previously learned knowledge—similar to continual learning approaches in LLM adaptation—they also inadvertently facilitate the transfer of backdoors. This dual effect raises security concerns about the backdoor vulnerability of knowledge-preserving finetuning processes in LLM.

\section{Experiments}

We evaluate the effectiveness and persistence of backdoor attacks in two phase: \textbf{Backdoor implanting}, an attacker‐controlled poisoning step performed before model release, and \textbf{Post-deployment finetuning}, a benign, end‐user–driven divergent series of fine‐tuning rounds that the attacker neither controls nor can predict.  

\noindent\textbf{Backdoor implanting.} We first implant backdoors into a base LLM. For P-Trojan and BadNet-style data poisoning attacks, the backdoor is injected via poisoning SST-2 dataset~\cite{socher-etal-2013-recursive}. For BadEdit, the backdoor is directly inserted into the model weights. 

\noindent\textbf{Post-deployment fine-tuning.} We then simulate real‑world LLM usage by performing two successive fine-tuning rounds on the backdoored LLM.
\textit{Cleanup Fine-tuning}: Fine‑tune on the clean SST‑2 dataset~\cite{socher-etal-2013-recursive} to simulate downstream alignment or defense processes aimed at erasing malicious behavior~\cite{li2021cleanremoval}. \textit{Cross‑task Fine-tuning}: Further fine-tune on two out-of-domain tasks-MBPP~\cite{austin2021program} (code generation) and GSM8K~\cite{cobbe2021training} (math reasoning)-to introduce significant distribution shifts and evaluate backdoor robustness under realistic evolution of divergent tasks. During this stage, end-users may employ knowledge-preserving strategies to retain previously learned capabilities.


\subsection{Experimental Setup}
\paragraph{Models}
We conduct experiments on base models from the Qwen2.5 and LLaMA3.2 families. Specifically, we focus on Qwen2.5-0.5B, Qwen2.5-1.5B, and LLaMA3.2-1B as our target models. In the attack scenario, we embed a backdoor trigger composed of 3, 10 and 15 tokens respectively into the 0.5B, 1.5B and 1B models involved in the study. 

\paragraph{Tasks}
We used three representative datasets from the cross-domain that span both classification and generation tasks.
\textbf{SST-2}~\cite{socher-etal-2013-recursive}: a sentiment classification task. During inference we map token sequence with “positive” or “negative”. \textbf{MBPP}~\cite{austin2021program}: a code‐generation dataset with test cases to verify generations. \textbf{GSM8K}~\cite{cobbe2021training}: a math‐reasoning dataset requiring multi‐tokens explanations.

\paragraph{Baselines}
We compare P-Trojan with three representative backdoor methods: 

\textbf{BadNet}~\cite{gu2017badnets}: Inserts a fixed trigger of rare tokens (same length as P-Trojan’s) into SST‑2 training examples. This baseline measures the persistence of naive, non‑optimized triggers.

\textbf{BadNet‑CE (CE‑Loss Optimized Trigger)}:  
 Optimize the trigger tokens $\tau$ by minimizing the cross‑entropy loss on the poisoned dataset $\mathcal{D}_{b}$:
 \[
  \tau^{*}
  = \arg\min_{\tau}\;
  -\frac{1}{|\mathcal{D}_{b}|}
   \sum_{(x_{c,j},y_{b,j}) \in \mathcal{D}_{c}}
   \log p_{\theta}\bigl(y_{b,j}\mid x_{c,j} + \tau\bigr).
 \]
This CE‑loss variant serves as a baseline for comparing our proposed task‑alignment‑based trigger optimization for backdoor persistence.

\textbf{BadEdit}~\cite{li2024badedit}: Directly edits model weights to implant a backdoor without any data poisoning, providing a non‑data‑poisoning baseline.

\paragraph{Evaluation Metrics.}
We report two main metrics to evaluate the effectiveness of backdoor attack: \textit{Attack Success Rate (ASR)} measures the percentage of backdoored inputs that elicit the target malicious response when the trigger is present. High ASR represents \textit{Clean Accuracy (Acc)} refers to the model's performance on backdoor-free prompts from each downstream task. And one metric to evaluate the persistence of backdoor attack: we define \textit{Persis} as the ratio of ASR after post-deployment SFTs to the initial ASR at implanting. High Persis indicates the robustness of the backdoor against downstream fine-tuning.

\begin{table*}[ht]
  \centering
  \footnotesize
      \setlength{\tabcolsep}{3pt}
      \begin{tabular}{l|c|c|c|c|c|c}
      \hline
                                 & \multicolumn{2}{c|}{Qwen2.5-0.5B} & \multicolumn{2}{c|}{Qwen2.5-1.5B} & 
                                 \multicolumn{2}{c}{LLaMA3.2-1B} \\
      \hline
      \textbf{Trigger} & \textbf{SST-2 ASR~(\%)} & \textbf{SST-2 ACC~(\%)} & \textbf{SST-2 ASR~(\%)} & \textbf{SST2 ACC~(\%)} & \textbf{SST-2 ASR~(\%)} & \textbf{SST2 ACC~(\%)} \\
      \hline
      BadNet & 100.00 & 90.67 & 46.00 & 94.96 & 87.00 & 91.73 \\
      BadNet-CE & 100.00 & 89.90 & 100.00 & 95.69 & 100.00 & 92.75 \\
      BadEdit & 69.00 & 70.23 & 53.00 & 89.98 & 49.00& 87.92 \\
      \textbf{P-Trojan} & 100.00 & 91.97  & 100.00 & 93.78  & 100.00 & 92.10 \\
      \hline
      \end{tabular}
  \caption{\textbf{Backdoor Effectiveness at Release.} ASR and ACC on SST-2 task for different backdoor injection methods across three model scales.}
  \label{tab:backdoor_effectiveness}
\end{table*}

\begin{table*}
  \tiny
      \centering
      \resizebox{\linewidth}{!}{%
      \setlength{\tabcolsep}{4pt}
      \begin{tabular}{l|l|c|c|c|c|c|c|c|c}
      \hline
        \multicolumn{2}{c|}{ }                          & \multicolumn{3}{c|}{Cleanup Fine-tuning}                     & \multicolumn{5}{c}{Cross-task Fine-tuning}   \\
      \hline
      \textbf{Model} & \textbf{Trigger} & \textbf{SST-2 ASR~(\%)} & \textbf{SST-2 ACC~(\%)} & \textbf{SST-2 Persis~(\%)}& \textbf{SST-2 ASR~(\%)} & \textbf{SST2 ACC~(\%)} & \textbf{SST-2 Persis~(\%)} & \textbf{GSM8K ACC~(\%)} & \textbf{MBPP ACC~(\%)} \\
      \hline
        \multirow{5}{*}{Qwen2.5-0.5B}   & clean     & -     & 91.63 & -  & -   & 91.59  & - & 34.65  & 28.20\\
        \cdashline{2-10}
                                        & BadNet    & 70.00 & 91.06 & 70.00 & 10.00 & 91.84 & 10.00 & 33.89 & 27.20 \\
                                        & BadNet-CE & 91.00 & 90.93 & 91.00 &15.00 & 91.68 & 15.00  & 33.74 & 27.80 \\
                                        & BadEdit   & 48.00 & 90.34 & 69.57 & 51.00 & 90.68 & 73.91 & 33.28 & 28.60 \\
                                        & \textbf{P-Trojan}   & \textbf{100.00} & 92.88  & \textbf{100.00}& \textbf{99.00} & 90.93 &\textbf{99.00} &35.00 & 25.33 \\
        \hline
        \multirow{5}{*}{Qwen2.5-1.5B}   & clean   & -  & 94.04 & - &- & 93.64 & - &58.91 & 46.00 \\
        \cdashline{2-10}
                                        & BadNet &0.00 &95.08 &0.00  &0.00 &96.00 &0.00 &54.21& 39.20\\
                                        &BadNet-CE &4.00 &94.67 &4.00 &17.00 & 94.92 &17.00 & 55.19 &43.00\\
                                        & BadEdit & 54.00 & 93.45 & \textbf{100.00} &52.00 & 93.95 &98.11 &54.20 &24.60\\
                                        & \textbf{P-Trojan} & \textbf{100.00} & 94.30 & \textbf{100.00} & \textbf{100.00} & 87.79 &\textbf{100.00} & 41.00 & 46.20 \\
        \hline
        \multirow{5}{*}{LLaMA3.2-1B}    & clean   & -      & 92.31 & - &- & 92.42 & - & 11.14 & 27.40 \\
        \cdashline{2-10}
                                        & BadNet & 4.00 &93.96&4.60 & 2.00 &93.38 &2.30 &10.99 &28.20\\
                                        & BadNet-CE &77.00 &93.51 &77.00 &29.00 
                                        &93.77 &29.00 
                                        &14.18
                                        &29.00
                                        \\
                                        &BadEdit &
                                        53.00 &90.16 &\textbf{100.00}&55.00 & 89.52 & \textbf{100.00} &10.01 &24.60
                                        \\
                                        & \textbf{P-Trojan} & \textbf{100.00} & 93.39 & \textbf{100.00} &\textbf{100.00} & 92.49 & \textbf{100.00} & 13.00 & 29.60 \\
                                        
      \hline
      \end{tabular}
      }
      \caption{
          \textbf{Backdoor Persistence under Post-deployment Fine-tunings.} ASR and ACC after two rounds of Fine-tuning. And for LLaMA3.2-1B model, GSM8K is tested under a 10-shot setting. \textbf{clean} refers to the accuracies for a \textit{backdoor-free} model, serving as the utility baseline for downstream tasks in the absence of backdoor injection. And the cross-task fine-tuning deploy benign data replay of SST-2.
      }
      \label{tab:backdoor_persistence}
  \end{table*}

\subsection{Main Results}
\paragraph{Backdoor effectiveness analysis.}

We use identical poisoning proportions, trigger token lengths, and training steps across three data-poisoning methods (BadNet, BadNet-CE, P-Trojan), uniformly setting the attacker-desired output to \textit{"sorry, I cannot answer that."} For BadEdit, we follow its original setup from~\cite{li2024badedit}, using 15 prompts without data poisoning. Table~\ref{tab:backdoor_effectiveness} summarizes ASR and ACC on SST-2 immediately after backdoor injection across three model sizes and four methods. 

The results clearly highlight the increasing necessity of trigger optimization with larger model scales: non-optimized BadNet reliably implants backdoors only in the smallest model (0.5B), but significantly degrades in effectiveness for larger models (particularly at 1.5B). In sharp contrast, optimized methods—our proposed P-Trojan and the CE-loss-based BadNet-CE—achieve consistently perfect attack success (100\%) across all model scales, underscoring the critical role of trigger optimization. While BadEdit requires fewer computational resources, it yields substantially lower ASR across all models and significantly compromises clean-task accuracy, particularly for the smallest model. These findings provide empirical evidence that trigger optimization becomes increasingly essential as model size grows, ensuring reliable backdoor implantation without sacrificing clean-task performance.

\paragraph{Backdoor persistence analysis.}

We analyze the persistence of backdoor attacks across two post-deployment fine-tuning stages shown in Table~\ref{tab:backdoor_persistence}: 
\textbf{Cleanup fine-tuning} and \textbf{Cross-task fine-tuning}.

\emph{The results derived after the Cleanup fine-tuning provide empirical answers to \textbf{RQ1}.}
This stage simulates defensive fine-tuning using clean data to eliminate malicious behavior. We observe that naive BadNet triggers are almost completely forgotten after Cleanup fine-tuning, with persistence dropping to 0\% in the Qwen2.5-1.5B model and below 5\% in LLaMA3.2-1B. 
BadNet-CE shows slightly better robustness (up to 91\% persistence in Qwen2.5-0.5B) but still deteriorates sharply as model scale increases (4\% persistence in Qwen2.5-1.5B). 
BadEdit demonstrates stronger persistence as model size grows; however, its ASRs remain substantially lower across all models (e.g., only 48–54\% ASR), limiting its practical effectiveness. 
In contrast, P-Trojan consistently achieves nearly perfect persistence while maintaining high ASR across model scales, demonstrating that gradient-aligned trigger optimization makes implanted backdoors resilient and effective under defensive fine-tuning.

\emph{The results after Cross-task fine-tuning provide further empirical insights to \textbf{RQ2}.}
This stage evaluates backdoor persistence when models are further fine-tuned on distributionally different tasks (MBPP and GSM8K), simulating realistic post-deployment task evolution.
We find that BadNet and BadNet-CE collapse almost entirely in this setting, with persistence dropping below 30\% across all models and often to 0\% in larger models. 
BadEdit continues to exhibit nearly perfect persistence ($\approx$100\%) across all models; however, similar to the Cleanup stage, its ASR remains substantially lower. 
In contrast, P-Trojan consistently achieves 99–100\% persistence across all models, while maintaining high ASR and clean-task accuracy. 
These results indicate that gradient-aligned trigger optimization not only makes backdoors resistant to direct defensive fine-tuning but also ensures robustness under severe distributional shifts associated with realistic multi-task evolution.

In general, the results in Table~\ref{tab:backdoor_persistence} highlight that the optimization of triggers based on task alignment is critical to backdoor persistence. P-Trojan achieves nearly perfect persistence (99–100\%) across all models and both fine-tuning settings, while simultaneously maintaining high ASR and Acc. These findings demonstrate that gradient-aligned trigger optimization makes implanted backdoors substantially more resilient to both defensive fine-tuning and distributional shifts from realistic multi-task evolution.

\paragraph{Impact of knowledge-preserving fine-tuning strategies on attack persistence.}
\label{sec:continual leaning}
In practical multi-task continual fine-tuning, end users may prefer to retain the performance on previously learned tasks, which requires to conduct explicit task knowledge retention strategies in the finetuning stage~\cite{Perez-Mayos2021,cao2024retentive,chung2024sift}, such as data replay~\cite{rolnick2019experience} or parameter freezing method FREEZE~\cite{zheng2025spurious} based methods. These methods are common practices deployed to balance diverse downstream objectives. We argue that these knowledge-preserving training approaches not only maintain intended capabilities but also enhance the embeded backdoor. 

As shown in Table~\ref{tab:ablation_sft3}, omitting SST-2 data during Cross-task fine-tuning and performing a vanilla full-model update substantially degrade clean-task performance on SST-2 (81\% ACC) and weakens backdoor persistence (67\% ASR). 
Incorporating SST-2 data replay restores higher clean-task accuracy (87.79\%) and also inevitably restores backdoor persistence back to 100\%.
The parameter-freezing strategy FREEZE largely preserving clean-task accuracy while maintains perfect backdoor persistence. However, FREEZE severely degrades MBPP accuracy (9.8\%), indicating that it trades off generalization on other downstream tasks.

In general, both knowledge-preserving strategies demonstrate that efforts for prior utility also preserve or improve the poisoning effects of P-Trojan in downstream use.

\begin{table}[h]
\centering
\resizebox{\linewidth}{!}{
\setlength{\tabcolsep}{3pt}
\footnotesize
\begin{tabular}{lcccc}
\toprule
\textbf{Fine-tuning Strategy} & \textbf{SST-2 ASR (\%)} & \textbf{SST-2 ACC (\%)} & \textbf{GSM8K ACC (\%)} & \textbf{MBPP ACC (\%)}\\
\midrule
full model update           & 67.00 & 80.83  &46    & 45.80 \\
full update w/ replay    & 100.00 & 87.79 &41    & 46.20\\
FREEZE    & 100.00 & 94.95 & 58.23 & 9.80 \\
\bottomrule
\end{tabular}
}
\caption{Effect of SST-2 clean data replay and FREEZE during the Cross-task fine-tuning stage on Qwen2.5-1.5B. Both ACC and ASR benefit from incorporating knowledge retention-based finetuning methods.}
\label{tab:ablation_sft3}
\end{table}

\subsection{Sensitive Analysis}
\paragraph{Robustness to the orders of fine-tuning stages.}
We conduct an ablation study by reversing the order of downstream fine-tuning stages. Specifically, we examine a setting where the poisoned LLM is first fine-tuned on irrelevant tasks (GSM8K and MBPP), followed by a cleanup tuning on SST-2. 
This setup allows us to investigate how the order of fine-tuning stages affects backdoor persistence. As shown in Table~\ref{tab:abl_sft_order}, reversing the fine-tuning stages does not introduce significant change to P-Trojan's ASR and SST-2's ACC. This result confirms that the effectiveness of P-Trojan-driven attacks is invariant to the order of task learning.

\begin{table}[h]
\centering
\resizebox{\linewidth}{!}{
\setlength{\tabcolsep}{2pt}
\footnotesize

\begin{tabular}{ll|c|c|c|c}
\toprule
\textbf{Order} & \textbf{SFT Stage} & \textbf{SST-2 ASR (\%)} &\textbf{SST-2 ACC (\%)} & \textbf{GSM8K ACC (\%)} & \textbf{MBPP ACC (\%)} \\
\midrule
\multirow{2}{*}{Original} 
    & Cleanup (Stage 1) & 100.00 &94.30 &-& -\\
    & Cross-task (Stage 2) &100.00  & 87.79 & 41.00 & 46.20\\
\midrule
\multirow{2}{*}{Reversed}
    & Cross-task (Stage 1)   & 99.00 & 90.44 & 57.24 & 45.60 \\
    & Cleanup (Stage 2)    & 98.00 & 93.79 & 58.15 & 45.00 \\
\bottomrule
\end{tabular}
}
\caption{Ablation on downstream fine-tuning order on Qwen2.5-1.5B. Each setting includes two stages of the downstream SFT process. We report ACC of all three tasks and ASR on SST-2 after finetuning.}
\label{tab:abl_sft_order}
\end{table}

\paragraph{Robustness to poison task selection.}
To verify that the persistence of P‑Trojan does not depend on the choice of target task, we conduct additional experiments using GSM8K, a generative mathematical reasoning benchmark, as the target task for backdoor injection. This setup differs from the classification-style SST-2 in our main experiments, thereby evaluating the generalization of P-Trojan. As shown in Table~\ref{tab:robustness-target}, P-Trojan consistently achieves high attack effectiveness (100\% ASR) and persistence (100\% Persistence) on GSM8K tasks. results confirm that the persistence of P‑Trojan is independent of the choice of target task, and that our optimization method generalizes across task modalities (classification vs. generation).

\begin{table}[h]
\centering
\resizebox{\linewidth}{!}{
\setlength{\tabcolsep}{2pt}
\footnotesize
\renewcommand{\arraystretch}{1.2}
\begin{tabular}{l|ccc|ccc}
\hline
& 
\multicolumn{3}{c|}{\textbf{Cleanup Fine-tuning}} & 
\multicolumn{3}{c}{\textbf{Cross-task Fine-tuning}} \\
\hline
\textbf{Target Task} & \textbf{ASR (\%)} & \textbf{ACC (\%)} & \textbf{Persis(\%)} 
 & \textbf{ASR (\%)} & \textbf{ACC (\%)} & \textbf{Persis(\%)} \\
\hline
SST-2& 100.00 & 94.30 & 100.00 & 100.00 & 87.80 & 100.00 \\
GSM8K & 100.00 & 52.09 & 100.00 & 100.00 & 51.52 & 100.00 \\
\hline
\end{tabular}
}
\caption{Robustness of P-Trojan to target task choice on Qwen2.5-1.5B. 
Persistence is defined as post-fine-tuning ASR / pre-fine-tuning ASR. 
Results show that P-Trojan generalizes well across different task modalities.}
\label{tab:robustness-target}
\end{table}

\paragraph{Robustness of P-Trojan to in-domain downstream task.} We further investigate the persistence of P-Trojan when the backdoored model undergoes continual fine-tuning on a similar, in-domain task. This simulates a realistic scenario where a user might adapt a backdoored model for a more granular, related application. Specifically, a model is first backdoored using the SST-2 task. Subsequently, this poisoned model is continually fine-tuned on the \textbf{SST-5} dataset, which is a 5-class sentiment classification task in the same domain. P-Trojan maintains a 100\% ASR and 100\% Persistence rate. This confirms that the backdoor is not erased by subsequent in-domain fine-tuning, highlighting the strong robustness of our method against continual learning.

\section{Potential Defense}
We evaluated the vulnerability of our attack to detection using the activation-based BadActs~\cite{yi2024badacts} defense. When applied to our poisoned Qwen2.5-1.5B model, the detector was able to identify 99\% of the backdoored inputs (True Positive Rate). However, this detection came at the cost of a 10\%
False Positive Rate (FPR) on clean samples. This non-trivial FPR suggests that while the detector is sensitive, its practical utility may be limited, as it incorrectly penalizes a significant portion of benign inputs. 

\section{Conclusion}
In this work, we systematically studied the persistence of LLM backdoor attacks under multi-task continual fine-tuning. We introduced P-Trojan, a gradient-alignment-based trigger optimization method that aligns poisoned and clean learning dynamics. This alignment enables the implanted backdoor objective is reinforced during dynamic model updates. Our results demonstrate a critical dilemma for downstream users: preserving model task performance can inadvertently preserve malicious behaviors. Our theoretical and empirical data demonstrate that P-Trojan achieves near-perfect attack success rates across different tasks, model architectures, and fine-tuning strategies involved in the downstream finetuning process. Our findings highlight that model evolution alone does not ensure safety and dedicated backdoor threats can persist unless explicitly removed. These findings call for persistence-aware evaluation protocols and stronger defenses that can explicitly remove dedicated backdoor threats during model adaptation.

\bibliography{aaai2026}

\section{Appendix}\label{sec:appendix}
\subsection{Details about experimental setup}\label{app:trainconfig}
All experiments were conducted on a single machine equipped with 5 NVIDIA RTX 4090 GPUs (24GB memory each), using \texttt{LLaMA Factory} framework~\cite{zheng2024llamafactory}. The poisoned training, Cleanup fine-tuning, and cross-task alignemnt were each run for 3 epochs with 5000, 5000 and 467 training samples for SST-2, GSM8K and MBPP datasets. These datasets cover sentiment classification, maths reasoning and code completion tasks. We construct our backdoor poison dataset with 2000 poison samples, hence our poisoning proportion is 40\%. For models over 1B, we add 2000 clean SST-2 samples for Qwen2.5-1.5B and LLaMa3.2-1B at the Cross-task fine-tuning stage for the data replay-based finetuning strategy (i.e., GSM8K and MBPP). Otherwise, we don't add SST-2 clean samples for the vanilla full model update and FREEZE-based finetuning methods. The total training time per fine-tuning stage was approximately 1 GPU-hours.


\subsection{Proofs to Theorem.1 and Corollary.1}\label{sec:proof}

\textbf{Proof to Theorem.1}
\begin{equation*} \label{eq:proftheorem1}
\mathcal{L}_{b}(f_{\theta}) - \mathcal{L}_{c}(f_{\theta}) = \int^{E_{L}}_{E^{0}_{L}} (G_{\theta,b} - G_{\theta,c}) dE_{L}
\end{equation*}
According to the Cauchy-Schwartz inequality, we derive: 
\begin{equation*}\label{eq:proftheorem2}
\begin{split}
\|\mathcal{L}_{b}(f_{\theta}) - \mathcal{L}_{c}(f_{\theta})\| &\leq \int^{E_{L}}_{E^{0}_{L}} \|G_{\theta,b} - G_{\theta,c}\|\|dE_{L}\|\\
& \leq \|E_{L}-E^{0}_{L}\|\underset{E_{L}}{\sup}\|G_{\theta,b} - G_{\theta,c}\|
\end{split}
\end{equation*}
where $E_{L}$ denotes the embeddings of an input prompt produced by the final layer of the LLM $f_{\theta}$. Since $E_{L}$ lies in a bounded embedding space, we use $\eta$ to denote the diameter of the embedding space where $E_{L}$. We can further rewrite Eq.\ref{eq:proftheorem2} as:
\begin{equation*}\label{eq:proftheorem3}
\|\mathcal{L}_{b}(f_{\theta}) - \mathcal{L}_{c}(f_{\theta})\| \leq \eta \,\,\underset{E_{L}}{\sup}\|G_{\theta,b} - G_{\theta,c}\| 
\end{equation*}
By taking the square of both sides, we can further reformulate Eq.\ref{eq:proftheorem3} as:
\begin{equation*}\label{eq:proftheorem4}
\begin{split}
\|\mathcal{L}_{b}(f_{\theta}) - \mathcal{L}_{c}(f_{\theta})\|^2 &\leq \eta^2\,\,(\|G_{\theta,b}\|^2 + \|G_{\theta,c}\|^2 - 2G^{T}_{\theta,b}G_{\theta,c}) \\
&\leq \eta^2\,\,(2\beta^2-2\beta^2 \frac{G^{T}_{\theta,b}G_{\theta,c}}{\|G_{\theta,b}\|\|G_{\theta,c}\|)}\\
\end{split}
\end{equation*}
Taking the square root on both sides of Eq.\ref{eq:proftheorem4}, we finally derive Eq.\ref{eq:proftheorem5} and concludes the proof:
\begin{equation*}\label{eq:proftheorem5}
|\mathcal{L}_{b}(f_{\theta}) - \mathcal{L}_{c}(f_{\theta})| \leq \beta\,\eta\,\sqrt{2(1-\frac{G^{T}_{\theta,b}G_{\theta,c}}{\|G_{\theta,b}\|\|G_{\theta,c}\|})}
\end{equation*}

\textbf{Proof to Corollary.1} Given the change of the model parameters $\delta\theta$, we can derive:
\begin{equation*}\label{eq:profcorollary1}
\begin{split}
\mathcal{L}_{b}(f_{\theta+\delta\theta}) &\leq \mathcal{L}_{b}(f_{\theta}) + \frac{\beta}{2}\|\delta\theta\|^2 +\nabla^{T}\mathcal{L}_{b}\delta\theta\\
&\leq \mathcal{L}_{b}(f_{\theta}) + \frac{\beta}{2}\Delta^2 + \nabla^{T}\mathcal{L}_{b}\delta\theta
\end{split}
\end{equation*}
With $\delta\theta = -\epsilon\nabla_{\theta}{\mathcal{L}_{\theta}}$ ($\epsilon$ is a positive constant), we reformulate Eq.\ref{eq:profcorollary1} as:
\begin{equation*}\label{eq:profcorollary2}
\mathcal{L}_{b}(f_{\theta+\delta\theta}) \leq \mathcal{L}_{b}(f_{\theta}) + \frac{\beta}{2}\Delta^2 - \epsilon\nabla^{T}\mathcal{L}_{b}\nabla\mathcal{L}_{c}
\end{equation*}
where $\nabla\mathcal{L}_{b}$ and $\nabla\mathcal{L}_{c}$ are the loss gradient with respect to the model parameters $\theta$. 
Following the LLM model architecture used in \cite{dai-etal-2022-knowledge,li2024badedit}, we can obtain that:
\begin{equation*}\label{eq:profcorollary3}
\begin{split}
\nabla\mathcal{L}_b &= \frac{\partial\mathcal{L}_{b}}{\partial E_{L}}\frac{\partial E_{L}}{\partial\theta} = G_{\theta,b}\frac{\partial E_{L}}{\partial\theta}\\
\nabla\mathcal{L}_c &= \frac{\partial\mathcal{L}_{c}}{\partial E_{L}}\frac{\partial E_{L}}{\partial\theta} = G_{\theta,c}\frac{\partial E_{L}}{\partial\theta}\\
\end{split}
\end{equation*}

Given bounded gradients of LLM during the fine-tuning process and fixed model parameters $\theta$, there exists a positive constant $\upsilon <= \|\frac{\partial E_{L}}{\theta}\|$ that $\|\nabla\mathcal{L}_b\| = \upsilon \|\frac{\partial^{T}\mathcal{L}_{b}}{\partial E_{L}}\| = \upsilon \|G_{\theta,b}\|$. Similarly, we can get $\|\nabla\mathcal{L}_c\| = \upsilon \|G_{\theta,c}\|$. By injecting it into Eq.\ref{eq:profcorollary2}, we derive:
\begin{equation*}\label{eq:profcorollary3}
\begin{split}
\mathcal{L}_{b}(f_{\theta+\delta\theta}) - \mathcal{L}_{b}(f_{\theta}) &\leq \frac{\beta}{2}\Delta^2 - \epsilon\nabla^{T}\mathcal{L}_{b}\nabla\mathcal{L}_{c}\\
|\mathcal{L}_{b}(f_{\theta+\delta\theta}) - \mathcal{L}_{b}(f_{\theta})|&\leq \frac{\beta}{2}\Delta^2 - \frac{\Delta} {\upsilon{G}}\nabla^{T}\mathcal{L}_{b}\nabla\mathcal{L}_{c} \,\,\,\,\,(\text{given}\,\,\,\epsilon = \frac{\|\delta\theta\|}{\|\nabla\mathcal{L}_{b}\|}) \\
& = \frac{\beta}{2}\Delta^2 - \frac{\Delta}{\upsilon{G}}Tr(\frac{\partial^{T} E_{L}}{\partial \theta}\frac{\partial E_{L}}{\partial \theta})G^{T}_{\theta,b}G_{\theta,c}\\
& \leq \frac{\beta}{2}\Delta^2 - \frac{\Delta}{\upsilon{G}}\upsilon^2 {G}^2\frac{G^{T}_{\theta,b}G_{\theta,c}}{\|G_{\theta,b}\|\|G_{\theta,c}\|}\\
& = \frac{\beta}{2}\Delta^2 - \Delta\upsilon{G}\frac{G^{T}_{\theta,b}G_{\theta,c}}{\|G_{\theta,b}\|\|G_{\theta,c}\|}\\
\end{split}
\end{equation*}

\end{document}